\documentclass[]{aa}
\usepackage{psfig}
\usepackage{txfonts}
\begin{document}

\title{Star formation in molecular cores III. \\
The effect of the turbulent power spectrum.}

\titlerunning{Star formation in molecular cores III}

\author{Simon\,P.\,Goodwin, A.\,P.\,Whitworth \and D.\,Ward-Thompson}

\authorrunning{S.\,P.\,Goodwin, A.\,P.\,Whitworth \and D.\,Ward-Thompson}

\offprints{Simon.Goodwin@astro.cf.ac.uk}

\institute{Dept. of Physics \& Astronomy, Cardiff University, 5 The 
Parade, Cardiff, CF24 3YB, Wales, UK}

\date{7/5/03}

%-------------------------------------------------------%
\abstract{
We investigate the effect of the turbulent power spectrum ($P(k) \propto 
k^{-n}$, with $n = 3,\,4\;{\rm or}\;5$) on the fragmentation of low-mass 
cores, by means of SPH simulations. We adopt initial density profiles and 
low levels of turbulence based on observation, and for each $n$-value we 
conduct an ensemble of simulations with different initial seeds for the 
turbulent velocity field, so as to obtain reasonable statistics. We find 
that when power is concentrated at larger scales (i.e. for larger $n$), 
more protostellar objects form and there is a higher proportion of low-mass 
stars and brown dwarfs. This is in direct contrast with the recent results 
of Delgado Donate et al., presumably because they adopted much higher 
levels of turbulence. 

}

\maketitle

%%%%%%%%%%%%%%%%%%%%%%%%%%%%%%%%%%%%%%%%%%%%%%%%%%%%%%%%%%%%%%%%%%%%%%%%%

\section{Introduction}

Stars form in dense molecular cores (eg. Andr\'e et al. 2000), and most 
stars -- especially young stars -- are in multiple systems (e.g. Duquennoy 
\& Mayor 1991; Mathieu 1994; Duch\^ene 1999, Patience et al. 2002), which 
implies that star-forming cores usually fragment into multiple objects.  

Simulations suggest that cores are prone to fragment into multiple 
objects under a variety of circumstances: (i) if they possess a 
small amount of initial rotation (eg. Burkert \& 
Bodenheimer 1996); (ii) if their collapse is triggered by a sudden 
increase in external pressure (eg. Hennebelle et al. 2003, 2004); or (iii) 
if they contain turbulence. The level of turbulence can be high (eg. 
Bate et al. 2002, 2003; Delgado Donate et al. 2004) or low (eg. Goodwin 
et al. 2004a,b).

This is the third in a series of papers investigating the collapse and 
fragmentation of cores with initial density profiles and levels of 
turbulence based on observation. In Paper I (Goodwin et al. 2004a) we 
have shown that cores with even a very low level of turbulence can 
fragment into multiple objects, but the number of fragments that form 
is very sensitive to the details of the initial turbulent velocity field. 
In paper II (Goodwin et al. 2004b) we have shown that the number of 
fragments increases as the level of turbulence is increased. In this paper 
we investigate the effect of the power spectrum of turbulence on the 
fragmentation of cores with {\it low} levels of turbulence.  A similar 
investigation has already been made by Delgado Donate et al. (2004) for 
cores with {\it high} levels of turbulence, and there are significant 
differences between their results and ours, which we explain in 
Section \ref{SECT:DISC}.

In Section 2 we describe the initial conditions and numerical methods 
used. In Section 3 we present our results, in Section 4 we discuss 
them, and in Section 5 we summarise our main conclusions.

%%%%%%%%%%%%%%%%%%%%%%%%%%%%%%%%%%%%%%%%%%%%%%%%%%%%%%%%%%%%%%%%%%%%%%%%%

\section{Initial conditions and numerical method}

The density profiles of prestellar cores are approximately flat in the 
centre, and then decrease as $r^{-\nu}$ with $2 \leq \nu \leq 5$ in their 
outer parts, until they merge with the background (e.g. Ward-Thompson et 
al. 1994, 1999; Andr\'e et al. 1996, 2000; Tafalla et al. 2004; Kirk et 
al. 2005).  A good fit to the density profile is given by 
\begin{equation} \label{EQN:DENSITY}
\rho(r) = \frac{\rho_{\rm kernel}}{(1 + (r/R_{\rm kernel})^2)^2} \,,
\end{equation}
where $\rho_{\rm kernel}$ is the central density and $R_{\rm kernel}$ is 
the radius of the region in which the density is approximately uniform
(cf. Whitworth \& Ward-Thompson 2001).  We set 
$\rho_{\rm kernel} = 3 \times 10^{-18}\,{\rm g}\,{\rm cm}^{-3}$ 
and $R_{\rm kernel} = 5,000\,{\rm AU}$, with the outer boundary of the 
core at $R_{\rm core} = 50,000\,{\rm AU}$, so the total mass of the core 
is $M_{\rm core} = 5.4\,{\rm M}_{\odot}$. These parameters are typical 
of low-mass star forming cores (eg. Jijina et al. 1999). The core is 
initially isothermal, with $T = 10\,{\rm K}$, and molecular, with mean 
gas-particle mass $\bar{m} = 4 \times 10^{-24}\,{\rm g}$. Hence the 
core has a ratio of thermal to gravitational energy of 
\begin{equation}
\alpha_{\rm therm} \equiv \frac{U_{\rm therm}}{|\Omega|} \simeq 0.3 \;.
\end{equation}

%%%%%%%%%%%%%%%%%%%%%%%%%%%%%%% FIGURE %%%%%%%%%%%%%%%%%%%%%%%%%%%%%%%%%%%%%%
\begin{figure*}
\centerline{\psfig{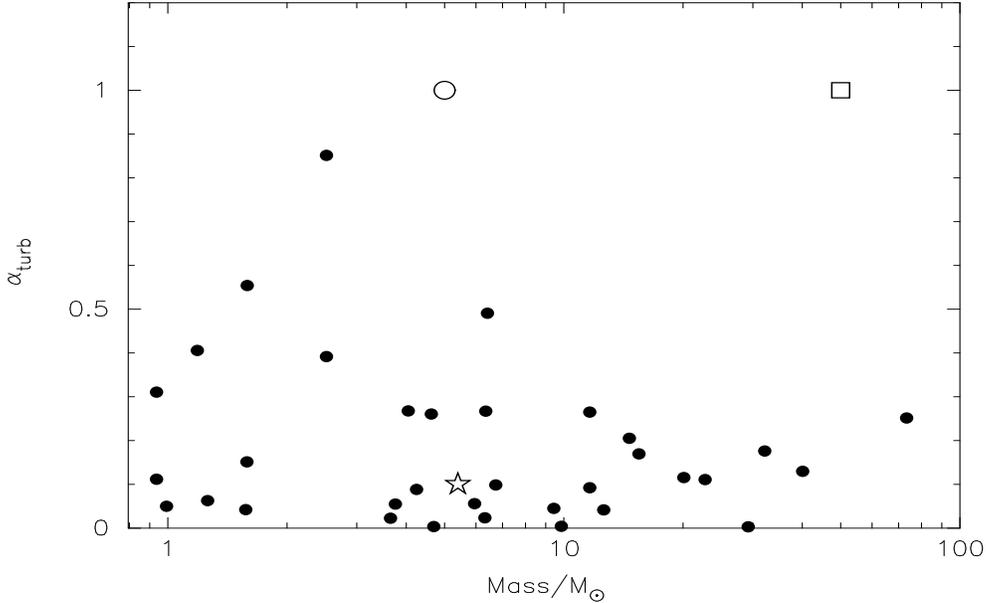}}
\caption{The filled circles give estimated values of $\alpha_{\rm
turb}$  (the ratio of turbulent to gravitational energy) and $M_{\rm
core}$ (core  mass) for the starless cores in the Jijina et al. (1999)
catalogue. The open  star shows the values used in this paper:
$\alpha_{\rm turb} = 0.1$ and  $M_{\rm core} = 5.4 M_\odot$. The open
circle shows the values used by  Delgado Donate et al. (2004):
$\alpha_{\rm turb} = 1.0$ and $M_{\rm core} =  5.0 M_\odot$. The open
square shows the values used by Bate et al. (2002,  2003):
$\alpha_{\rm turb} = 1.0$ and $M_{\rm core} = 50 M_\odot$. 
Some  of the cores in the Jijina et al. catalogue have already started
to collapse,  and therefore systematic infall motions are making a
contribution to their  non-thermal line-widths, thus causing us to
over-estimate their $\alpha_{\rm turb}$.  Hence, we believe
that our choice of $\alpha_{\rm turb}$ is more representative of the
initial conditions in these cores.}
\label{fig:nonthermal}
\end{figure*}
%%%%%%%%%%%%%%%%%%%%%%%%%%%%%%%%%%%%%%%%%%%%%%%%%%%%%%%%%%%%%%%%%%%%%%%%%%%%%%

The line widths of molecular cores show a significant non-thermal
contribution (eg. Myers 1983; Myers et al. 1991; Jijina et al. 1999),
which is attributable to internal
turbulence. Fig.~\ref{fig:nonthermal}  shows the estimated ratios of
turbulent to gravitational energy,
\begin{equation}
\alpha_{\rm turb} \equiv \frac{U_{\rm turb}}{|\Omega|} \,,
\end{equation}
and the estimated masses, $M_{\rm core}$, for prestellar cores from
the Jijina  et al. (1999) catalogue. These cores have been selected as
prestellar on the basis  of having low temperature ($<20\,{\rm K}$),
no associated IRAS source and no  observed outflow. Using
Fig.~\ref{fig:nonthermal} as a guide, we adopt  $\alpha_{\rm turb} =
0.1$, and hence a total virial ratio of $\alpha_{\rm turb}
+\alpha_{\rm therm} \sim 0.4$, for all the simulations in this paper.

To model the turbulence in a core, we impose a divergence-free
gaussian random  velocity field with power spectrum $P(k) \propto
k^{-n}$ (cf. Bate et al. 2002,  2003; Fisher 2004; Bonnell et
al. 2003; Delgado-Donate et al. 2003,  2004).  The observed velocity
fields in GMCs and cores are well represented by  turbulent power
spectra of this form with $n=3$ to $4$ (Burkert \& Bodenheimer  2000).

In this paper we present two ensembles of 10 simulations each, one
with $n=3$  and one with $n=5$, and compare these with the ensemble of
20 simulations with  $n=4$ already presented in Paper II.  Within an
ensemble, each simulation  differs only in the random seed used to
initialise the turbulent  velocity field.

\subsection{Computation method and constitutive physics}

The simulations are performed with the {\sc dragon} code (Goodwin et
al. 2004a),  which is based on a standard implementation of SPH
(eg. Monaghan 1992). An  octal tree (Barnes \& Hut 1984) is used to
evaluate gravitational accelerations  and to identify SPH
neighbours. Particle smoothing lengths are adjusted so that  each
particle has $N_{\rm neib} = 50 \pm 5$ neighbours. Gravity is kernel
softened with the particle smoothing lengths, and standard artificial
viscosity  is included with $\alpha_v=1$ and $\beta_v=2$. The
interested reader is referred  to Paper I for further details.

At low densities, radiative cooling is efficient and the gas in a core is 
approximately isothermal at $T_{_0} \sim 10\,{\rm K}$. However, once the 
density exceeds $\rho_{\rm crit} \sim 10^{-13}\,{\rm g}\,{\rm cm}^{-3}$, 
the optical depth through a core becomes too large for efficient cooling 
(Larson 1969; Tohline 1982; Masunaga \& Inutsuka 2000) and the gas switches 
to being approximately adiabatic. We model this behaviour with a barotropic 
equation of state (eg. Tohline 1982; Masunaga \& Inutsuka 2000): 
\begin{equation}
\frac{P}{\rho} = c_{_0}^2 \left[ 1 + \left
( \frac{\rho}{\rho_{\rm crit}} \right)^{2/3} \right] \,.
\end{equation}
Here $P$ is the pressure, $\rho$ is the density, and $c_{_0} \approx 0.19\, 
{\rm km}\,{\rm s}^{-1}$ is the isothermal sound speed in molecular gas at 
$T \simeq 10\,{\rm K}$.

Following the evolution of dense regions that are collapsing to stellar 
densities is very expensive computationally. In order to avoid this expense, 
wherever a bound regions forms with $\rho > 100\,\rho_{\rm crit}$ we replace 
it with a sink particle (Bate et al. 1995). A sink particle interacts with 
the gas gravitationally, and it accretes any SPH particle which (a) approaches 
closer than $10\,{\rm AU}$ and (b) is bound to it.   When an SPH
particle is accreted by a sink particle, the sink particle acquires the
mass, linear momentum and angular momentum of the SPH particle, and
therefore we can still monitor the conservation of these quantities as
a check on the fidelity of the code.  We refer to sink particles 
generically as `objects'; and then, more specifically, as `stars' when the 
sink mass is greater than $0.08 {\rm M}_{\odot}$, and as `brown dwarfs' when 
the mass is lower than this.

%%%%%%%%%%%%%%%%%%%% SUMMARY TABLE 1 %%%%%%%%%%%%%%%%%%%%%%%%%%%%%%%%%%%%%%%%
\begin{table*}
\caption[]{For each value of $n$, we list the number of realisations simulated 
(${\cal N}_{\rm real}$); the mean mass in objects at the end of a simulation 
($M_{\rm tot}/M_\odot$); the mean number of objects formed per simulation 
$\left (\bar{\cal N}_{\rm obj} \right)$ and its variance; the mean number 
of stars ejected from a core (${\cal N}_{\rm ej}$); the net ratio of brown 
dwarfs to stars (${\cal N}_{\rm BD}/{\cal N}_*$); the fraction of all 
objects that are single ($S/{\cal N}_{\rm obj}$); the median
semi-major axis ($a_{\rm med}$); the mean mass-ratio for 
binary systems ($\bar{q}$); the variance of the mass ratio of binary
systems ($\sigma_q$); and the mean numbers of singles ($\bar{S}$), 
binaries ($\bar{B}$), triples ($\bar{T}$), quadruples ($\bar{Q}$), and 
quintuples ($\bar{Q}'$) formed by one core.}
\label{tab:summary}
\begin{center}
\begin{tabular}{ccccccccccccccc}
$n$ & ${\cal N}_{\rm real}$ & $M_{\rm tot}/M_\odot$ 
 & $\bar{\cal N}_{\rm obj}$ & ${\cal N}_{\rm ej}$ 
 & ${\cal N}_{\rm BD}/{\cal N}_*$ & $S/{\cal N}_{\rm obj}$ 
 & ${a_{\rm med}} $ & $\bar{q}$ & $\sigma_q$ & $\bar{S}$ & $\bar{B}$ & 
$\bar{T}$ & $\bar{Q}$ & $\bar{Q}'$ \\ \hline
3  & 10 & 3.39 & $3.7\pm1.4$ & 0.80 & 0.08 & 0.22 & 13 & 0.74 & 0.17 &
 0.80 & 0.20 & 0.30 & 0.40 & 0 \\
4  & 20 & 3.35 & $4.8\pm3.1$ & 1.95 & 0.20 & 0.41 & 9  & 0.83 & 0.26 &
 1.95 & 0.25 & 0.35 & 0.25 & 0.05 \\
5  & 10 & 3.41 & $5.5\pm3.0$ & 2.10 & 0.15 & 0.38 & 6  & 0.65 & 0.21 &
 2.10 & 0.20 & 0.40 & 0.20 & 0.20 \\ \hline
\end{tabular}
\end{center}
\end{table*}
%%%%%%%%%%%%%%%%%%%%%%%%%%%%%%%%%%%%%%%%%%%%%%%%%%%%%%%%%%%%%%%%%%%%%%%%%

%%%%%%%%%%%%%%%%%%%%%%%%%%%%%%%%%%%%%%%%%%%%%%%%%%%%%%%%%%%%%%%%%%%%%%

\section{Results}

We have performed two ensembles of 10 simulations each, one ensemble using 
$P(k) \propto k^{-3}$ and one ensemble using $P(k) \propto k^{-5}$. These 
are then compared with the ensemble of 20 simulations using $P(k) \propto 
k^{-4}$ reported in Paper II. All simulations treat cores with 
$\alpha_{\rm turb} = 0.10$. A summary of the results is given in 
Table~\ref{tab:summary}.  The details of each simulation are presented 
in Table~\ref{tab:runs}.

%%%%%%%%%%%%%%%%%%%%%%%%%%%%%%%%%%%%%%%%%%%%%%%%%%%%%%%%%%%%%%%%%%%%%%%%

\subsection{The number of objects formed}

The fragmentation of a collapsing, mildly turbulent core usually 
starts with the formation of a primary protostar surrounded by a 
rotating, but very unrelaxed, disc. The inflow of material onto 
the disc is very inhomogeneous and irregular, and as it joins the 
disc it causes spiral arms to develop. If these arms become 
sufficiently dense, they fragment to form secondary companions 
(e.g. Goodwin et al. 2004a,b; Gawryszczak et al. 2006). If $n$ is 
low, most of the turbulent energy is concentrated on small scales, 
but the resulting inhomogeneities in the inflow are of such low 
amplitude that the spiral perturbations they seed in the disc 
tend to be dissipated by the shear in the disc rather than being 
amplified by self-gravity. Conversely, if $n$ is large, most of 
the turbulent energy is concentrated on large scales, and although 
the resulting inhomogeneities are again of low amplitude, they are 
of sufficiently large mass that the spiral perturbations they seed 
in the disc have a better chance of being amplified by self-gravity 
and fragmenting into secondary companions.  As $n$ is increased 
from $n = 3$ to $n = 5$, more power is invested in 
large-scale turbulence, and therefore there tend to be more objects 
formed. When $n=3$, $\bar{\cal N}_{\rm obj} = 3.7 \pm 1.4$; but when 
$n=5$, this increases to $\bar{\cal N}_{\rm obj} = 5.5 \pm 3.0$.

We stress that the large variance on $\bar{\cal N}_{\rm obj}$ is because 
this is a chaotic process, and two simulations from the same ensemble (same 
$\alpha_{\rm turb}$ and $n$) can produce two vastly different sets of 
objects. For example one simulation from the ensemble with $n=4$ produces 
a quadruple, a binary and four singles, whilst several others produce just 
one star (see Table~\ref{tab:runs}).  Nonetheless, a generic
  trend is seen.

%%%%%%%%%%%%%%%%%%%%%%%%%%%%%%%%%%%%%%%%%%%%%%%%%%%%%%%%%%%%%%%%%%%%%%%

\subsection{Mass functions and companion probabilities}

Once an object forms, its final mass is determined by competitive accretion 
(Bonnell et al. 2001) and dynamical interaction with other objects. 
Competitive accretion causes the more massive objects, and/or those which 
reside in the dense material at the centre of the core, to grow rapidly in 
mass. Dynamical interaction causes some objects, usually the lower-mass ones, 
to be ejected from the dense material at the centre of the core, so that the 
remaining objects become more tightly bound and some eventually end up in stable 
multiple systems. Thus, in general, it is the lower-mass objects (low-mass 
stars and brown dwarfs) which are ejected as singles before they can 
accrete much mass (Reipurth \& Clarke 2001); and the higher-mass objects 
($\sim 1\,{\rm M}_\odot$) which remain in the centre of the core, and 
form multiples.

\begin{figure*}
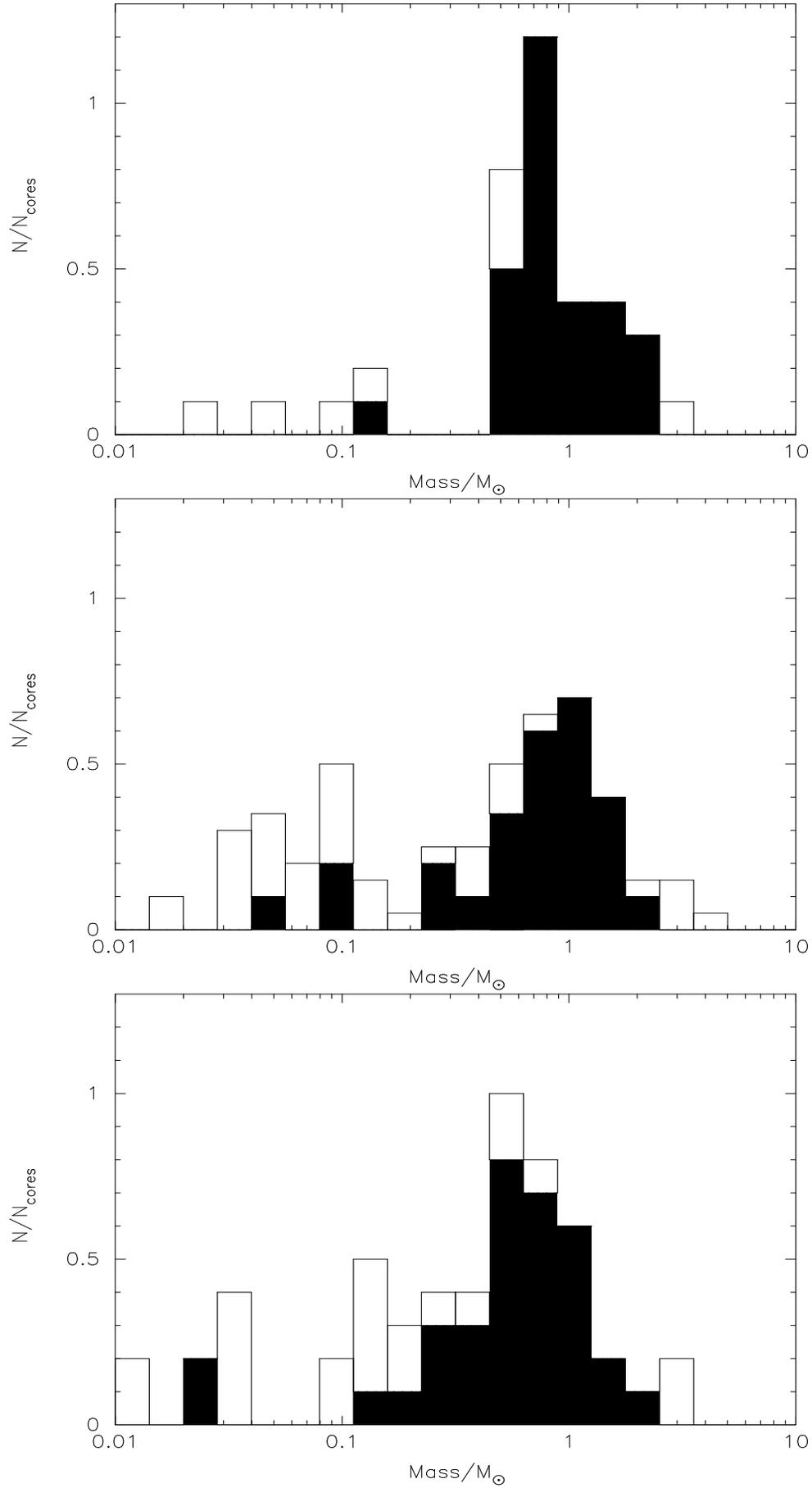

\centerline{\psfig{figure=4026fig2a.ps,height=8.0cm,width=13.0cm,angle=270}}
\centerline{\psfig{figure=4026fig2b.ps,height=8.0cm,width=13.0cm,angle=270}}
\centerline{\psfig{figure=4026fig2c.ps,height=8.0cm,width=13.0cm,angle=270}}
\caption{Normalised mass functions for objects spawned by cores with $n=3$ 
(top), $n=4$ (middle) and $n=5$ (bottom). The filled portion of each 
histogram represents objects in multiple systems, while the open portion 
represents single objects.}
\label{fig:masses}
\end{figure*}

Figure~\ref{fig:masses} shows the mass functions from the three ensembles 
with $n=3$ (top), $n=4$ (middle) and $n=5$ (bottom). In each case the 
filled portion of the histogram represents objects in multiple systems, 
and the open portion represents single objects.

The mass functions for $n=4$ and $n=5$ are very similar. There is a broad 
peak around $1\,{\rm M}_\odot$, consisting mainly of objects in multiple 
systems, and a flat tail of lower-mass objects consisting mainly of ejected 
singles. Because of the large number of ejected singles, the overall companion 
probability is low, $\sim 0.6$.

In contrast, when $n=3$, the mass function is dominated by the peak around 
$\sim 1\,{\rm M}_\odot$ and there are very few ejected singles. This is 
because, when $n=3$, an individual core spawns fewer objects, and therefore 
fewer ejections are required to stabilise the remaining multiple system. The 
paucity of ejected singles gives a much higher overall companion probability, 
$\sim 0.8$.

%%%%%%%%%%%%%%%%%%%% MASS FUNCTIONS FIGURE %%%%%%%%%%%%%%%%%%%%%%%%%%%%%%%
%%%%%%%%%%%%%%%%%%%%%%%%%%%%%%%%%%%%%%%%%%%%%%%%%%%%%%%%%%%%%%%%%%%%%%%%%%

%%%%%%%%%%%%%%%%%%%% MASS RATIOS FIGURE %%%%%%%%%%%%%%%%%%%%%%%%%%%%%%%%%
\begin{figure*}
\centerline{\psfig{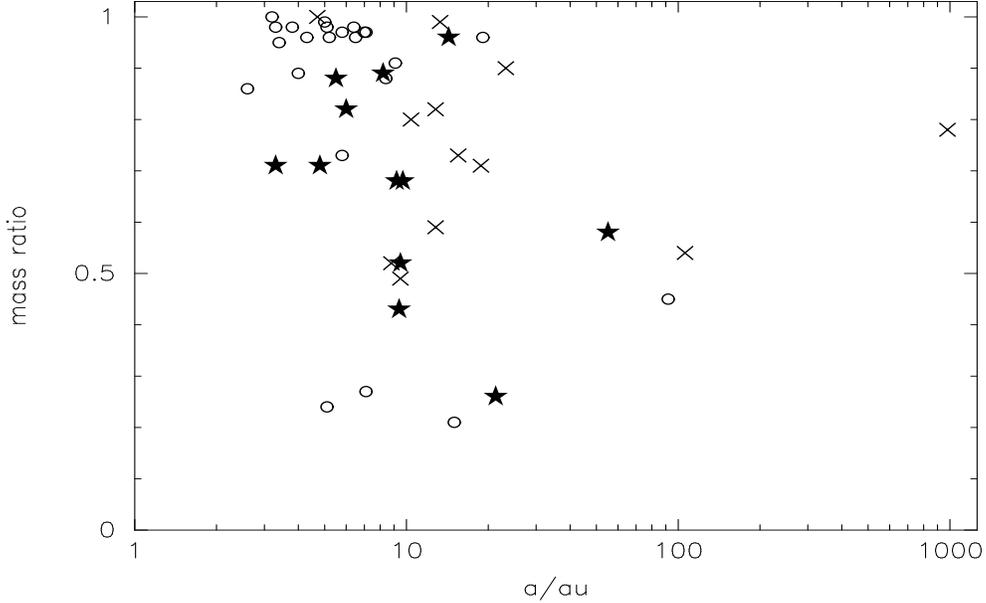}}
\caption{Mass ratio, $q$, against separation, $a$, for binaries spawned 
by cores with $n=3$ (crosses), $n=4$ (open circles) and $n=5$ (solid 
stars).}
\label{fig:massratios}
\end{figure*}
%%%%%%%%%%%%%%%%%%%%%%%%%%%%%%%%%%%%%%%%%%%%%%%%%%%%%%%%%%%%%%%%%%%%%%%%%%

%%%%%%%%%%%%%%%%%%%%% TABLE 2 %%%%%%%%%%%%%%%%%%%%%%%%%%%%%%%%%%%%%%%%%%
\begin{table*}
\caption[]{The results of the individual simulations; in all cases
$\alpha_{\rm turb} = 0.10$. Column 1 gives the simulation identifier and 
Column 2 gives $n$ (the exponent of the turbulent power spectrum, 
$P(k) \propto k^{-n}$). Column 3 gives $M_{\rm obj}$ (the total mass of 
objects formed, stars plus brown dwarfs), Column 4 gives 
${\cal N}_{\rm obj}$ (the total number of objects formed) and Column 5 
gives ${\cal N}_{\rm bd}$ (the total number of brown dwarfs formed). 
Column 6 gives the multiplicities of the multiple systems formed, and 
Column 7 gives the mass of each individual object. Those objects which 
are part of a binary system are distinguished with $^b$, those which 
are part of a triple system with $^t$, and those which are part of a 
quadruple or quintuple system with $^q$. Two realisations have ejected 
binary systems.}
\label{tab:runs}
\begin{tabular}{lllllll}
\hline
&&&&&\\
 {\rm ID} & $n$ & $M_{\rm obj}$ & $N_{\rm obj}$ &
$N_{\rm bd}$ & Multiplicity & Masses/$M_{\odot}$ \\
&&&&&\\
\hline
&&&&&\\

571 & 3 & 3.04 & 4  & 0 & {\rm Triple}     & 1.06$^t$, 0.77$^t$,
0.61, 0.59$^t$ \\
572 & 3 & 3.21 & 5  & 0 & {\rm Quadruple}  & 1.24$^q$, 0.75$^q$,
0.60$^q$, 0.47, 0.14$^q$ \\
573 & 3 & 3.62 & 5  & 1 & {\rm Quadruple}  & 1.43$^q$, 0.78$^q$,
0.75$^q$, 0.62$^q$, 0.04 \\
574 & 3 & 3.20 & 5  & 0 & {\rm Triple}     & 0.89$^t$, 0.80$^t$,
0.80$^t$, 0.57, 0.14 \\
575 & 3 & 2.95 & 1  & 0 & {\rm Single}     & 2.95 \\
576 & 3 & 3.79 & 5  & 1 & {\rm Quadruple}  & 1.05$^q$, 1.05$^q$,
0.86$^q$, 0.74$^q$, 0.09 \\
577 & 3 & 3.58 & 2  & 0 & {\rm Binary}     & 2.01$^b$, 1.57$^b$ \\
578 & 3 & 2.97 & 4  & 1 & {\rm Triple}     & 2.04$^t$, 0.45$^t$,
0.45$^t$, 0.02 \\
579 & 3 & 3.79 & 2  & 0 & {\rm Binary}     & 2.47$^b$, 1.33$^b$ \\
580 & 3 & 3.76 & 4  & 0 & {\rm Quadruple}  & 1.36$^q$, 0.81$^q$,
0.80$^q$, 0.79$^q$ \\

% ----------------------------------------------

001 & 4 & 3.78 & 3  & 0 & {\rm Triple}     & 1.49$^t$, 1.15$^t$,
1.13$^t$ \\
002 & 4 & 2.83 & 1  & 0 & {\rm Single}     & 2.38\\
003 & 4 & 3.72 & 1  & 0 & {\rm Single}     & 3.72 \\
004 & 4 & 3.48 & 1  & 0 & {\rm Single}     & 3.48 \\
005 & 4 & 2.86 & 4  & 1 & {\rm Binary}     & 1.43$^b$, 0.77, 0.65$^b$,
0.02 \\
006 & 4 & 2.84 & 1  & 0 & {\rm Single}     & 2.84 \\
007 & 4 & 3.15 & 5  & 0 & {\rm Triple\,\,\&\,\,Binary} & 1.76$^t$,
0.72$^t$, 0.47$^t$, 0.10$^b$, 0.10$^b$  \\
008 & 4 & 3.22 & 6  & 2 & {\rm Quadruple}  & 1.97$^q$, 0.47$^q$,
0.35$^q$, 0.34$^q$, 0.06, 0.03 \\
009 & 4 & 3.48 & 8  & 4 & {\rm Quadruple}  & 2.28$^q$, 0.49$^q$,
0.26$^q$, 0.25$^q$, 0.08, 0.05, 0.04, 0.04 \\
010 & 4 & 3.31 & 8  & 1 & {\rm Quadruple}  & 0.76$^q$, 0.74$^q$,
0.58$^q$, 0.57$^q$, 0.46, 0.09, 0.08, 0.03 \\
011 & 4 & 3.96 & 12 & 4 & {\rm Triple\,\,\&\,\,binary?} & 0.89$^t$, 
0.82$^t$, 0.82$^t$, 0.42, 0.38, 0.25,
0.12, 0.11, 0.04$^b$, 0.04$^b$, 0.03, 0.03  \\
012 & 4 & 3.60 & 6  & 2 & {\rm Triple}     & 1.34$^t$, 0.92$^t$,
0.79$^t$, 0.50, 0.04, 0.02\\
013 & 4 & 3.18 & 10 & 3 & {\rm Quadruple\,\,\&\,\,binary} & 0.77$^q$,
0.68$^q$, 0.61$^q$, 0.60$^q$, 0.11$^b$, 0.11$^b$, 0.10, 0.06, 0.05,
0.04 \\
014 & 4 & 3.29 & 4  & 1 & {\rm Binary}     & 1.58$^b$, 1.16$^b$, 0.49,
0.08 \\
015 & 4 & 2.48 & 1  & 0 & {\rm Single}     & 2.48 \\
016 & 4 & 3.58 & 4  & 0 & {\rm Triple}     & 1.23$^t$, 1.15$^t$,
1.11$^t$, 0.09 \\
017 & 4 & 3.41 & 8  & 0 & {\rm Quintuple}  & 1.10$^q$, 0.98$^q$,
0.32$^q$, 0.27$^q$, 0.27$^q$, 0.17, 0.15, 0.14  \\
018 & 4 & 3.48 & 4  & 0 & {\rm Quadruple}  & 0.98$^q$, 0.94$^q$,
0.79$^q$, 0.77$^q$ \\
019 & 4 & 3.58 & 5  & 1 & {\rm Triple}     & 1.38$^t$, 1.03$^t$,
1.00$^t$, 0.11, 0.06 \\
020 & 4 & 3.77 & 3  & 0 & {\rm Triple}     & 1.28$^t$, 1.27$^t$, 1.22$^t$\\

% ----------------------------------------------

551 & 5 & 3.21 & 1  & 0 & {\rm Single}     & 3.21 \\
552 & 5 & 3.34 & 7  & 0 & {\rm Quadruple}  & 0.69$^q$, 0.59$^q$,
0.57$^q$, 0.52$^q$, 0.45, 0.37, 0.15 \\
553 & 5 & 3.46 & 6  & 0 & {\rm Triple}     & 1.29$^t$, 1.11$^t$,
0.49, 0.29$^t$, 0.17, 0.10 \\
554 & 5 & 3.22 & 8  & 2 & {\rm Quadruple}  & 1.10$^q$, 0.75$^q$,
0.71$^q$, 0.27$^q$, 0.18, 0.15, 0.04, 0.01 \\
555 & 5 & 3.52 & 5  & 0 & {\rm Quintuple}  & 1.16$^q$, 0.84$^q$,
0.60$^q$, 0.51$^q$, 0.40$^q$ \\
556 & 5 & 3.39 & 1  & 0 & {\rm Single}     & 3.39 \\
557 & 5 & 3.77 & 7  & 1 & {\rm Triple + Binary} & 1.45$^t$, 0.93$^t$,
0.89$^t$, 0.22$^b$, 0.13$^b$, 0.13, 0.01 \\
558 & 5 & 3.10 & 11 & 3 & {\rm Quintuple + Binary} & 0.83$^q$,
0.57$^q$, 0.43$^q$, 0.35$^q$, 0.30$^q$, 0.29, 0.14, 0.09, 0.04, 
0.02$^b$, 0.02$^b$ \\
559 & 5 & 3.75 & 6  & 2 & {\rm Triple}     & 1.13$^t$, 0.97$^t$,
0.80$^t$, 0.79, 0.03, 0.03 \\
560 & 5 & 3.30 & 3  & 0 & {\rm Triple}     & 2.25$^t$, 0.59$^t$,
0.46$^t$ \\ \hline
\end{tabular}
\end{table*}
%%%%%%%%%%%%%%%%%%%%%%%%%%%%%%%%%%%%%%%%%%%%%%%%%%%%%%%%%%%%%%%%%%%%%%%%%%

\subsection{Binary separations and mass ratios}

In all cases ($n = 3,\,4\,{\rm and}\;5$), the distribution of separations 
is much narrower than that observed by Duquennoy \& Mayor for local G 
dwarfs, which is not surprising, since we have considered only one core 
mass, and only one level of turbulence\footnote{Hubber \& Whitworth (2005) 
have shown how the full range of binary parameters can be reproduced by 
considering a range of core parameters}. In particular there is a total lack of 
wide binaries ($a > 100\,{\rm AU}$). For $n=3$, most binaries have separations 
in the range 10 to $30\,{\rm AU}$, and there is only one hard binary ($a < 
10\,{\rm AU}$). In contrast, when $n=4$, 17 of the 20 binaries formed have 
$a < 10\,{\rm AU}$; and when $n=5$, 10 of the 13 binaries formed have $a < 
10\,{\rm AU}$. This difference arises because a core with larger $n$ tends to 
spawn more objects, and so on average more ejections are needed before a stable 
multiple is left; specifically, the average number of objects ejected from a 
core is only 0.80 for $n=3$, but 1.95 for $n=4$ and 2.10 for $n=5$. Since each 
ejection hardens the multiple that is left behind, the greater number of ejections 
for $n=4$ and $n=5$ means harder multiples, i.e. smaller separations. 

Fig.~\ref{fig:massratios} shows the distribution of mass ratio, $q \equiv 
M_{_2} / M_{_1}$, against separation, $a$ (strictly, semi-major axis). In 
all cases the mean mass ratio is high, viz. $\bar{q}=0.74,\,0.83
\,{\rm and}\;0.65$ for $n=3,\,4\,{\rm and}\;5$, respectively. There is a 
tendency for closer binaries to have higher mass-ratios, i.e. more nearly 
equal components. This tendency arises because the formation of a close 
binary often entails hardening by ejection, and the ejections tend to 
remove the less massive objects; hence only the more massive objects are 
left as potential binary components, and the range of possible masses is 
thereby reduced, pushing $q$ towards unity. In addition, when a close binary 
accretes material with high angular momentum, the accreted material tends to 
end up on the less massive component, which again pushes $q$ towards unity 
(Whitworth et al. 1995, Bate \& Bonnell 1997, Papers I \& II).

Such high mass ratios are not compatible with the observations,
which, whilst they show a trend to more equal-mass companions in
close binaries (e.g. Mazeh et al. 1992; White \& Ghez 2001; Fisher
et al. 2005), are certainly not as extreme as suggested by these
results.  However, a recent paper by Yasuhiro et al. (2005) suggests
that accretion in a proto-binary system is generally onto the primary
(especially in circular orbits) as angular momentum is removed from 
the accreting gas by spiral shocks.  Such a mechanism is beyond the 
ability of these simulations to resolve and may help solve this 
problem.

\section{Discussion} \label{SECT:DISC}

Delgardo Donate et al. (2004) have also explored the effect of the 
turbulent power spectrum on the fragmentation of low-mass cores, 
performing two ensembles of 5 SPH simulations each, one with $n=3$ and one 
with $n=5$. They use a slightly different core mass ($5.0\,{\rm M}_\odot$, 
as compared with our $5.4\,{\rm M}_\odot$), a different density profile 
(uniform, as compared with Eqn. \ref{EQN:DENSITY}), and a slightly 
different equation of state. However, the most significant difference is 
that their cores have a much higher initial level of turbulence; 
specifically, they adopt $\alpha_{\rm turb}= 1.0$, as compared with our 
$\alpha_{\rm turb} = 0.1$. As a consequence, the initial turbulent 
velocities in their cores are mildly supersonic and the cores are 
marginally unbound ($\alpha_{\rm turb} + \alpha_{\rm therm} \simeq 1.1$), 
whereas the initial turbulent velocities in our cores are subsonic and 
the cores are approximately virialised ($\alpha_{\rm turb} + 
\alpha_{\rm therm} \simeq 0.4$). The amount of turbulent energy has a 
profound influence on the outcome of collapse.

In the strongly turbulent cores of Delgado Donate et al. (2004), there 
is so much power in the turbulence that even the small-scale 
inhomogeneities created by small-scale turbulent motions can become 
self-gravitating and collapse (whereas in our simulations these small-scale 
inhomogeneities have much lower amplitude and tend to disperse). As a result, 
many more objects are formed, and -- {\it in direct contrast with our results} 
-- more objects are formed when $n=3$ than when $n=5$. This is because 
the same amount of power invested in small scales (large $k$) produces a 
larger number of inhomogeneities than when it is invested in large scales 
(small $k$). Furthermore, if more objects are formed in a core, then 
there have to be more ejections before a stable multiple is created, so 
the mass functions derived by Delgado Donate et al. (2004) have a larger 
tail of low-mass singles (low-mass stars and brown dwarfs), {\it and} 
this tail is larger when $n=3$ than when $n=5$.

With reference to Fig. \ref{fig:nonthermal}, we suggest that the level of 
turbulence we have adopted is more representative of the cores in the 
Jijina et al. (1999) catalogue, and therefore probably more representative 
of the initial conditions in low-mass star forming cores, particularly if, 
as seems likely, some of the Jijina et al. cores are already collapsing, 
and therefore some of their nonthermal linewidth is attributable to 
collapse rather than initial turbulence. The low levels of turbulence we 
have invoked also seem to result in more acceptable values for the overall 
multiplicity.

%%%%%%%%%%%%%%%%%%%%%%%%%%%%%%%%%%%%%%%%%%%%%%%%%%%%%%%%%%%%%%%%%%%%%%%%%

\section{Conclusions}

We have investigated the influence of the slope of the turbulent power 
spectrum on the fragmentation of dense molecular cores, by means of an 
ensemble of SPH simulations. We consider a spherical, $5.4M_\odot$ 
core, with a Plummer-like density profile (Eqn. \ref{EQN:DENSITY}), and a 
low level of turbulence, $\alpha_{\rm turb} \equiv E_{\rm turb}/|\Omega|=0.10$, 
similar to observed cores such as L1544. The turbulence has a power 
spectrum $P(k) \propto k^{-n}$ with $n=3,\,4,\,{\rm or}\;5$. The choice 
of $n$ influences the number of objects that form, the mass function of 
those objects, and the properties of the multiple systems that they 
comprise. However, the process is chaotic, in the sense that even if $n$ is 
fixed, different realizations of the turbulent velocity field can 
produce widely different stellar masses and binary properties. Our main 
conclusions are therefore statistical in nature, and could always be 
improved by performing more simulations:

\begin{itemize}

\item{The average number of objects that form in a collapsing core 
increases monotonically with $n$, from $\bar{\cal N}_{\rm obj} = 3.7 
\pm 1.4$ when $n=3$, to $\bar{\cal N}_{\rm obj} = 5.5 \pm 3.0$ when 
$n=5$.}

\item{The mass function always involves a peak at $M \sim 1 
{\rm M}_\odot$ and most of the objects in the peak are in multiple systems.}

\item{As $n$ increases and more objects are produced, more dynamical 
ejections are required before a stable multiple is formed, which has 
two consequences. (i) A larger proportion of single, low-mass objects 
(low-mass stars and brown dwarfs) is produced; hence the mass function 
develops a low-mass tail and the mean multiplicity decreases. (ii) The 
resulting binaries tend to be harder, i.e. to have smaller separations.}

\item{The mean mass ratios of binary systems do not depend strongly on 
$n$, but close binaries tend to have mass ratios closer to unity, i.e. 
more nearly equal components. This is because the ejections which harden 
a binary preferentially remove low-mass objects, leaving the two most 
massive objects; the larger the initial number of objects, the more 
ejections are required, the harder the final binary, and the closer the 
masses of the two binary components.}

\item{The low level of turbulence we have adopted in these simulations 
($\alpha_{\rm turb} = 0.1$) appears to be in good agreement with 
observations of low-mass cores, and to reproduce the mean multiplicity 
in observed stellar populations.}

\end{itemize}

These results do not match observed multiple systems very well.
The separation distribution is too narrow and the mass ratios tend too
much towards equal-masses.  This is probably due in part to the fact
that we consider only one (rather high) core mass.  The separation
distribution may well improve if a realistic range of core masses were
considered (cf. Hubber \& Whitworth 2005), but this would vastly
increase the computational load as a far larger series of ensembles
would be required.  Even then, the small separation range is exacerbated
by the hardening of systems through ejections when too many
fragments are formed (see Goodwin \& Kroupa 2005).  The tendency to
equal-mass binaries may be a result of unresolved physics in the inner
accretion region (see Yasuhiro et al. 2005).

%%%%%%%%%%%%%%%%%%%%%%%%%%%%%%%%%%%%%%%%%%%%%%%%%%%%%%%%%%%%%%%%%%%%%%%%%

\begin{acknowledgements}

SPG is a UKAFF Fellow.  We thank B. Sathyaprakash and R. Balasubramanian 
for allowing us to perform these simulations on their Beowulf
cluster.  Part of this work was carried out while DWT was on sabbatical at the 
Observatoire de Bordeaux, and he gratefully acknowledges the hospitality 
accorded to him there.

\end{acknowledgements}


\begin{thebibliography}{}

\bibitem[]{}Andr\'{e}, P., Ward-Thompson, D. \& Barsony, M. 2000, in 
'Protstars \& Planets IV', eds. V. Mannings, A. P. Boss \& S. S. 
Russell (University of Arizona Press: Tuscon), p 59

\bibitem[]{}Andr\'e, P., Ward-Thompson, D. \& Motte, F. 1996, 
A\&A, 314, 625

\bibitem[]{}Barnes, J. \& Hut, P. 1986, Nature, 324, 446

\bibitem[]{}Bate, M. R. \& Bonnell, I. A. 1997, MNRAS, 285, 33

\bibitem[]{}Bate, M. R., Bonnell, I. A. \& Bromm, V. 2002, MNRAS, 336, 705

\bibitem[]{}Bate, M. R., Bonnell, I. A. \& Bromm, V. 2003, MNRAS, 339, 577

\bibitem[]{}Bate, M. R., Bonnell, I. A. \& Price, N. M. 1995, MNRAS,
277, 362

\bibitem[]{}Bonnell, I. A., Bate, M. R., Clarke, C. J. \& Pringle, J. E. 
2001, MNRAS, 323, 785

\bibitem[]{}Bonnell, I. A., Bate, M. R. \& Vine, S. G. 2003, MNRAS,
343, 413

\bibitem[]{}Bonnell, I. A. 2003, private communication

\bibitem[]{}Burkert, A. \& Bodenheimer, P. 2000, ApJ, 543, 822

\bibitem[]{}Delgardo-Donate, E. J., Clarke, C. J. \& Bate, M. R. 2003,
MNRAS, 342, 926

\bibitem[]{}Delgardo-Donate, E. J., Clarke, C. J. \& Bate, M. R. 2004,
MNRAS, 347, 759

\bibitem[]{}Duch\^{e}ne, G. 1999, A\&A, 341, 547

\bibitem[]{}Duquennoy, A. \& Mayor, M. 1991, A\&A, 248, 485

\bibitem[]{}Fisher, R. T. 2004, ApJ, 600, 769

\bibitem[]{}Fisher, J., Schr\"oder, K-P. \& Smith, R. C. 2005, MNRAS,
  361, 495

\bibitem[]{}Gawryszczak, A. J., Goodwin, S. P., Burkert, A. \&
  R\'o\.zyczka, M. 2006, submitted to A\&A

\bibitem[]{}Goodwin, S. P. \& Kroupa, P. 2005, A\&A, 439, 565

\bibitem[]{}Goodwin, S. P., Whitworth, A. P. \& Ward-Thompson,
D. 2004a, A\&A, 414, 633 (Paper I)

\bibitem[]{}Goodwin, S. P., Whitworth, A. P. \& Ward-Thompson,
D. 2004b, A\&A, 423, 169 (Paper II)

\bibitem[]{}Hennebelle, P., Whitworth, A. P., Gladwin, P. P. \& 
Andr\'e, P. 2003, MNRAS, 340, 870 

\bibitem[]{}Hennebelle, P., Whitworth, A. P., Cha, S.-H. \& Goodwin,
S. P. 2004, MNRAS, 348, 687

\bibitem[]{}Hubber, D. A. \& Whitworth, A. P. 2005, A\&A, in press (astrop-ph/0503412)

\bibitem[]{}Jijina, J., Myers, P.C. \& Adams, F.C. 1999, ApJS, 125,
161

\bibitem[]{}Kirk, J. M., Ward-Thompson, D. \& Andr\'e, P. 2005, MN 
in press (astro-ph/0505190)

\bibitem[]{}Larson, R. B. 1969, MNRAS, 145, 271

\bibitem[]{}Masunaga, H. \& Inutsuka, S. 2000, ApJ, 531, 350

\bibitem[]{}Mathieu, R. D. 1994, ARA\&A, 32, 465 

\bibitem[]{}Mazeh, T., Goldberg, D., Duquennoy, A. \& Mayor, M. 1992, ApJ, 401, 265

\bibitem[]{}Monaghan, J. J. 1992, ARA\&A, 30, 543

\bibitem[]{}Myers, P. C. 1983, ApJ, 270, 105

\bibitem[]{}Myers, P. C., Ladd, E. F. \& Fuller, G. A. 1991, ApJ, 372, L95

\bibitem[]{}Patience, J., Ghez, A. M., Reid, I. N. \& Matthews, K. 2002, 
AJ, 123, 1570

\bibitem[]{}Reipurth, B. \& Clarke, C. 2001, ApJ, 122, 432

\bibitem[]{}Reipurth, B. \& Zinnecker, H. 1993, A\&A, 278, 81

\bibitem[]{}Tafalla, M., Myers, P. C., Caselli, P. \& Walmsley, C. M. 
2004, A\&A, 416, 191

\bibitem[]{}Tohline, J. E. 1982, Fundamentals of Cosmic Physics, 8, 1

\bibitem[]{}Ward-Thompson, D., Motte, F. \& Andr\'{e}, P. 1999, MNRAS, 305, 143

\bibitem[]{}Ward-Thompson, D., Scott, P. F., Hills, R. E. \& Andr\'e, 
P, 1994, MNRAS, 268, 276

\bibitem[]{}White, R. J. \& Ghez, A. M. 2001, ApJ, 556, 265

\bibitem[]{}Whitworth, A. P., Chapman, S. J., Bhattal, A. S., Disney, M. J., 
Pongracic, H. \& Turner, J. A. 1995, MNRAS, 277, 727

\bibitem[]{}Whitworth, A. P. \& Ward-Thompson, D. 2001, ApJ, 547, 317

\bibitem[]{}Yasuhiro, O., Sugimoto, K. \& Hanawa, T. 2005, ApJ, 623, 922

\end{thebibliography}
\end{document}